\documentclass{svproc}
\usepackage{graphicx}
\usepackage{booktabs}
\usepackage{url}

\title{Predicting MMSE Score from Finger-Tapping Measurement}
\titlerunning{Predicting MMSE Score} 
\author{Jian Ma}
\authorrunning{J. Ma} 
\institute{Hitachi (China) Research \& Development Corporation, Beijing, China \\  \email{majian@hitachi.cn}} 

\begin{document}
\maketitle

\begin{abstract}
Dementia is a leading cause of diseases for the elderly. Early diagnosis is very important for the elderly living with dementias. In this paper, we propose a method for dementia diagnosis by predicting MMSE score from finger-tapping measurement with machine learning pipeline. Based on measurement of finger tapping movement, the pipeline is first to select finger-tapping attributes with copula entropy and then to predict MMSE score from the selected attributes with predictive models. Experiments on real world data show that the predictive models such developed present good prediction performance. As a byproduct, the associations between certain finger-tapping attributes (`Number of taps', `Average of intervals', and `Frequency of taps' of both hands of bimanual in-phase task) and MMSE score are discovered with copula entropy, which may be interpreted as the biological relationship between cognitive ability and motor ability and therefore makes the predictive models explainable. The selected finger-tapping attributes can be considered as dementia biomarkers.
\end{abstract}

\section{Introduction}

The world population is aging \cite{un2017} and the diseases spectrum of world changes accordingly. In 2013, the WHO study \cite{who2013} estimated the top 20 leading causes of death in 2030 showing that noncommunicable diseases, instead of communicable disease, will become the major threats to human lives. Among them are dementias, which are common in the elderly people. In the World Alzheimer Report 2009 \cite{prince2009world}, Alzheimer’s Disease International estimated that 36 million people worldwide were living with dementia, with numbers doubling every 20 years to 66 million by 2030, and 115 million by 2050. 

Early diagnosis is crucial for dementia patients to make a timely plan of treatment, intervention and long-term disease management, which will definitely benefit not only patients themselves but their families, caregivers and health service providers. However, research \cite{prince2018world} showed that most patients currently living with dementia have not received a timely formal diagnosis. In high income countries, only 20-50\% of dementia cases are diagnosed and documented in primary care, and this gap between early diagnosis and unidentified patients is even much greater in low and middle income countries.

Diagnosing dementia involves cognitive assessment which measures brain functions, such as memory, thinking, language skills, attention, problem-solving, and many other mental abilities \cite{petersen2001practice}. Additionally, diagnosis based on blood tests and brain imaging tests may also be helpful for checking for biological evidences to find or rule out the cause of certain symptoms in clinical practice.

Since the first criteria for clinical diagnosis of Alzheimer Disease (also called the NINCDS-ADRDA criteria) was proposed at 1984 \cite{mckhann1984clinical}, clinical guideline for AD diagnosis has been updated several times (at 1994 \cite{1994PracticePF}, 2001 \cite{doody2001practice}, 2007 \cite{dubois2007research}, 2011 \cite{albert2011the}). In those updates criteria for blood tests and brain imaging tests were evolving as new pathophysiological evidences were accumulated, while criteria for cognitive tests has experienced enduring clinical successes and hardly been revised. Among many candidate instruments for cognitive assessment, the MMSE (Mini-Mental State Examination) has always been recommended as the preference for diagnosis of dementia due to its reliable performance in practice \cite{mckhann1984clinical,1994PracticePF,doody2001practice,dubois2007research,albert2011the}.

The MMSE, proposed by Folstein et al. \cite{folstein1975mini}, is a 30-points instrument for cognitive assessment, consists of 7 groups of questions measuring different aspects of mental state, and takes about 4-20 minutes to conduct. It has been reported to enjoy a very popularity among clinicians since its birth \cite{tombaugh1992the}.

Though there are many reliable instruments for cognitive assessment besides MMSE (such as Diagnostic and Statistical Manual, Kokmen Short Test, among many others), those tests are still considered to be too complicated and time-consuming for certain clinical settings. In this sense, simple instrument with less time cost and comparable performance is dearly needed for early diagnosis of dementia. Previously, researchers have developed a type of finger-tapping device \cite{kandori2004quantitative}, tried to utilize it as diagnosis tool to estimate MMSE score, and achieved promising results \cite{suzumura2016assessment}. In this research, we will propose a method for predicting MMSE score from finger-tapping attributes developed with Machine Learning (ML) pipeline.

The contributions of this paper are as follows:
\begin{enumerate}
	\item A ML pipeline comprising of Copula Entropy (CE) based variable selection and predictive models is proposed, which is generally applicable to other similar ML problems and can lead to explainable ML models;
	\item The associations between certain attributes of finger-tapping movement and MMSE score are discovered by CE, which may be interpreted as the relationship between cognitive and motor ability of the elderly people;
	\item A method for predicting MMSE score from the selected finger-tapping attributes, is proposed and its effectiveness is validated on real world data. Due to the above associations, the predictive models are explainable for clinical use.
\end{enumerate}

This paper is orgnized as follows: the methodology is presented in Section \ref{s:method}; Section \ref{s:exp} presents experiments and results; some discussion is given in Section \ref{s:discussion}; Section \ref{s:con} concludes the paper.

\section{Methodology}
\label{s:method}
\subsection{Copula Entropy}
\label{s:CopEnt}
\subsubsection{Theory}
Copula theory is about the representation of multivariate dependence with copula function \cite{joe2014dependence,nelsen1998an}. At the core of copula theory is Sklar theorem \cite{sklar1959fonctions} which states that multivariate probability density function can be represented as a product of its marginals and copula density function which represents dependence structure among random variables. Such representation seperates dependence structure, i.e., copula function, with the properties of individual variables -- marginals, which make it possible to deal with dependence structure only regardless of joint distribution and marginal distribution. This section is to define an statistical independence measure with copula. For clarity, please refer to \cite{ma2011mutual} for notations.

With copula density, Copula Entropy is define as follows \cite{ma2011mutual}:
\begin{definition}[Copula Entropy]
	\label{d:ce}
	Let $\mathbf{X}$ be random variables with marginal distributions $\mathbf{u}$ and copula density $c(\mathbf{u})$. CE of $\mathbf{X}$ is defined as
	\begin{equation}
	H_c(\mathbf{X})=-\int_{\mathbf{u}}{c(\mathbf{u})\log{c(\mathbf{u})}}d\mathbf{u}.
	\end{equation}
\end{definition}

In information theory, MI and entropy are two different concepts \cite{cover1991elements}. In \cite{ma2011mutual}, Ma and Sun proved that they are essentially same -- MI is also a kind of entropy, negative CE, which is stated as follows: 
\begin{theorem}
	\label{thm1}
	MI of random variables is equivalent to negative CE:
	\begin{equation}
	I(\mathbf{X})=-H_c(\mathbf{X}).
	\end{equation}
\end{theorem}
\noindent
The proof of Theorem \ref{thm1} is simple \cite{ma2011mutual}. There is also an instant corollary (Corollary \ref{c:ce}) on the relationship between information of joint probability density function, marginal density function and copula density function.
\begin{corollary}
	\label{c:ce}
	\begin{equation}
	H(\mathbf{X})=\sum_{i}{H(X_i)}+H_c(\mathbf{X}).
	\end{equation}
\end{corollary}
The above results cast insight into the relationship between entropy, MI, and copula through CE, and therefore build a bridge between information theory and copula theory. CE itself provides a mathematical theory of statistical independence measure.

\subsubsection{Estimation}
\label{s:est}
It has been widely considered that estimating MI is notoriously difficult. Under the blessing of Theorem \ref{thm1}, Ma and Sun \cite{ma2011mutual} proposed a simple and elegant non-parametric method for estimating CE (MI) from data which comprises of only two steps\footnote{The \textsf{R} package \textsf{copent} for estimating CE is available on CRAN and also on GitHub at \url{https://github.com/majianthu/copent}.}:
\begin{enumerate}
	\item Estimating Empirical Copula Density (ECD);
	\item Estimating CE.
\end{enumerate}

For Step 1, if given data samples $\{\mathbf{x}_1,\ldots,\mathbf{x}_T\}$ i.i.d. generated from random variables $\mathbf{X}=\{x_1,\ldots,x_N\}^T$, one can easily estimate ECD as follows:
\begin{equation}
F_i(x_i)=\frac{1}{T}\sum_{t=1}^{T}{\chi(\mathbf{x}_{t}^{i}\leq x_i)},
\end{equation}
where $i=1,\ldots,N$ and $\chi$ represents for indicator function. Let $\mathbf{u}=[F_1,\ldots,F_N]$, and then one can derive a new samples set $\{\mathbf{u}_1,\ldots,\mathbf{u}_T\}$ as data from ECD $c(\mathbf{u})$. In practice, Step 1 can be easily implemented non-parametrically with rank statistic.

Once ECD is estimated, Step 2 is essentially a problem of entropy estimation which has been contributed with many existing methods. Among them, the kNN method \cite{kraskov2004estimating} was suggested in \cite{ma2011mutual}. With rank statistic and the kNN method, one can derive a non-parametric method of estimating CE, which can be applied to any situation without any assumption on the underlying system.

\subsection{Predictive Models}
In this paper, two types of ML models, i.e., Linear Regression (LR) and Support Vector Regression (SVR), are selected among many others for building predictive models since LR is the most typical linear model and SVR is the most widely-used nonlinear model for small sample cases.

LR models linear relationship between dependent and some independent random variables. Suppose there are dependent random variable Y and an independent random vector X, the LR model is as:
\begin{equation}
	y = \mathbf{A}x+\beta + \varepsilon
\end{equation}
where $\mathbf{A},\beta$ are parameters to be estimated, and $\varepsilon$ is noise. 

SVR is a popular ML method that learns complex relationship from data \cite{smola2004a}. Theoretically, SVR can learn the model with simple model complexity and meanwhile do not compromise on predictive ability, due to the max-margin principle. The learning of SVR model is formulated as an optimization problem \cite{smola2004a}, which can be solved by quadratic programming techniques after transformed to its dual form. SVR has its nonlinear version with kernel tricks. The final SVR model is represented as
\begin{equation}
	f(x) = \sum_{i}{v_i k(x,x_i)+b}
\end{equation}
where $x_i$ represents support vector, and $k(\cdot,\cdot)$ represents kernel function.

\subsection{The Machine Learning pipeline}
\label{s:pipeline}
We propose a ML pipeline for predicting MMSE score with the above concept and methods. The pipeline starts from the finger-tapping attributes generated from the collected raw data (magnetic data of finger-tapping movement). With CE as association measure, the pipeline then selects the attributes mostly associated with MMSE score. Next, such selected attributes are fed into the trained predictive models (LR and SVR) to predict MMSE scores. The predictive model in the pipeline is not limited to the above mentioned two models, and open to others models. The advantage of this ML pipeline over other methods has been demonstrated on the UCI heart disease data \cite{jian2019variable}.

\section{Experiments and Results}
\label{s:exp}
\subsection{Data}
The data in the experiments were collected at Tianjin and Beijing with finger-tapping device \cite{kandori2004quantitative}. To collect data from finger-tapping test, two type of movements were measured for each test: bimanual in-phase, bimanual unti-phase, which derives 168 attributes totally from both hands on aspects of distance, velocity, acceleration, time, etc. (For more details on the attributes, please refer to \cite{suzumura2016assessment}). In the experiments, each movement lasts for 15 seconds. All the participants signed informed consent. At Tianjin, 40 people were recruited as subjects, whose age range at 45-84. From 20 out of 40 subjects, finger-tapping data were collected, the rest subjects only presented corrupted or noisy data. A total 142 finger-tapping tests were performed by these 20 subjects. At Beijing, 117 people were recruited as subjects, whose age range at 48-96. Each subject presented one sample data. Each subject at both locations was tested to derive a MMSE score after each finger-tapping task was performed. Note that subjects at Tianjin are mostly healthy people and present high MMSE scores while those at Beijing are mostly with dementias and present relatively low MMSE scores. 

\subsection{Experiments}
With the above data, we did experiments on predicting MMSE score. Experiments used the ML pipeline mentioned in Section \ref{s:pipeline}. Association between the finger-tapping attributes and MMSE was measured with CE and the most associated attributes were selected as inputs to the predictive models. The predictive models were trained and then test on a testing dataset. To evaluate the selected associated attributes on prediction performance, several additional attributes with less association strength were included as inputs of the predictive models in another experiment. In the experiments, CE was estimated with the non-parametric method in Section \ref{s:est} implemented in the \textsf{R} package \texttt{copent}\cite{copent}, and SVR is implemented in the \textsf{R} package \texttt{e1071} \cite{Chang2011LIBSVMAL}, the hyper-parameters of which were tuned to get optimal prediction results. 

For prediction, the whole dataset was randomly separated into two parts, 80\% for training the predictive models and 20\% for testing the accuracy of the trained  models. The Mean-Absolute-Error (MAE) was used to measure the performance of the predictive model on testing data: 
\begin{equation}
	MAE = \frac{1}{T}{\sum_{t=1}^{T}|S_{t}^{pred} - S_{t}^{true}|}
\end{equation}
where $S_{t}^{pred}$ and $S_{t}^{true}$ represent for predicted and true MMSE scores respectively. The final MAE is an average on the results of 100 run of experiments with the above setting. To further measure the power of the proposed method, we divided the samples with clinical cutoff value (MMSE=24) into healthy people and patient, and then calculated the diagnosis accuracy of the method.

\subsection{Results}
\begin{figure}	
\centering
\includegraphics[width=12 cm]{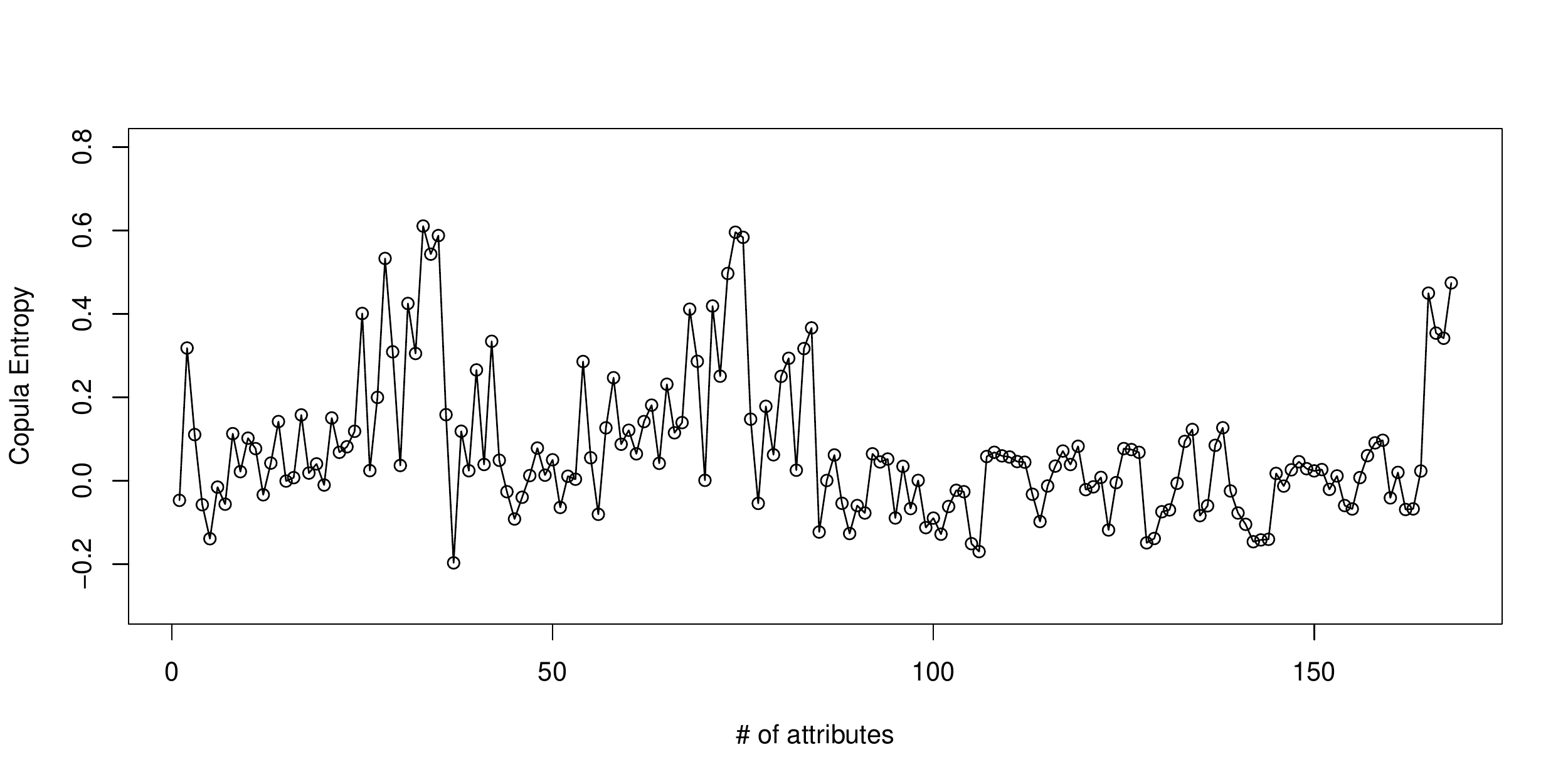}
\caption{Associations between MMSE and finger-tapping attributes.}
\label{fttjmi1}
\end{figure}

\begin{figure}
	\centering
	\includegraphics[width=8 cm]{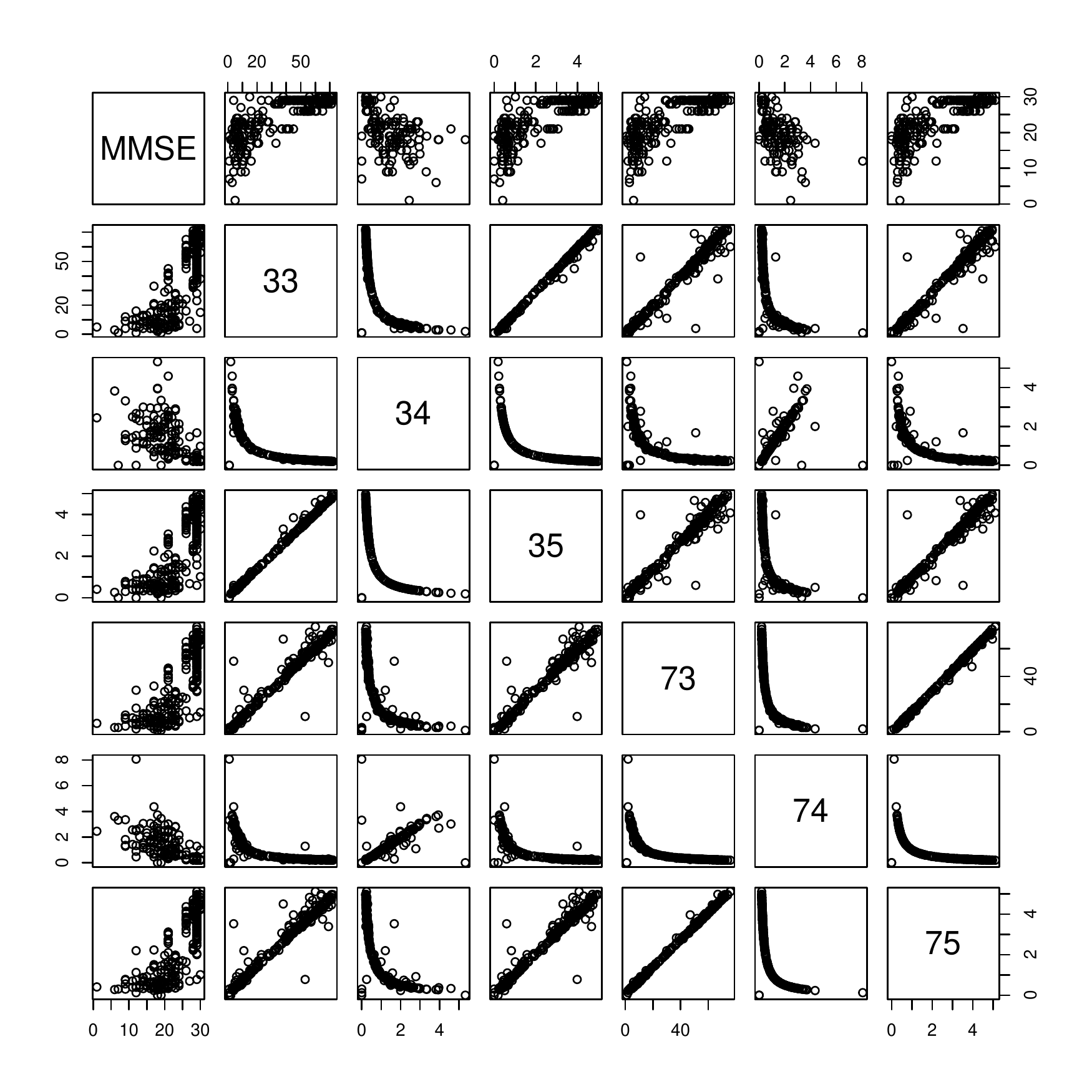}
	\caption{Scattering plot of MMSE and the associated attributes.}
	\label{pairs}
\end{figure}

\begin{figure}
\centering
\includegraphics[width=8 cm]{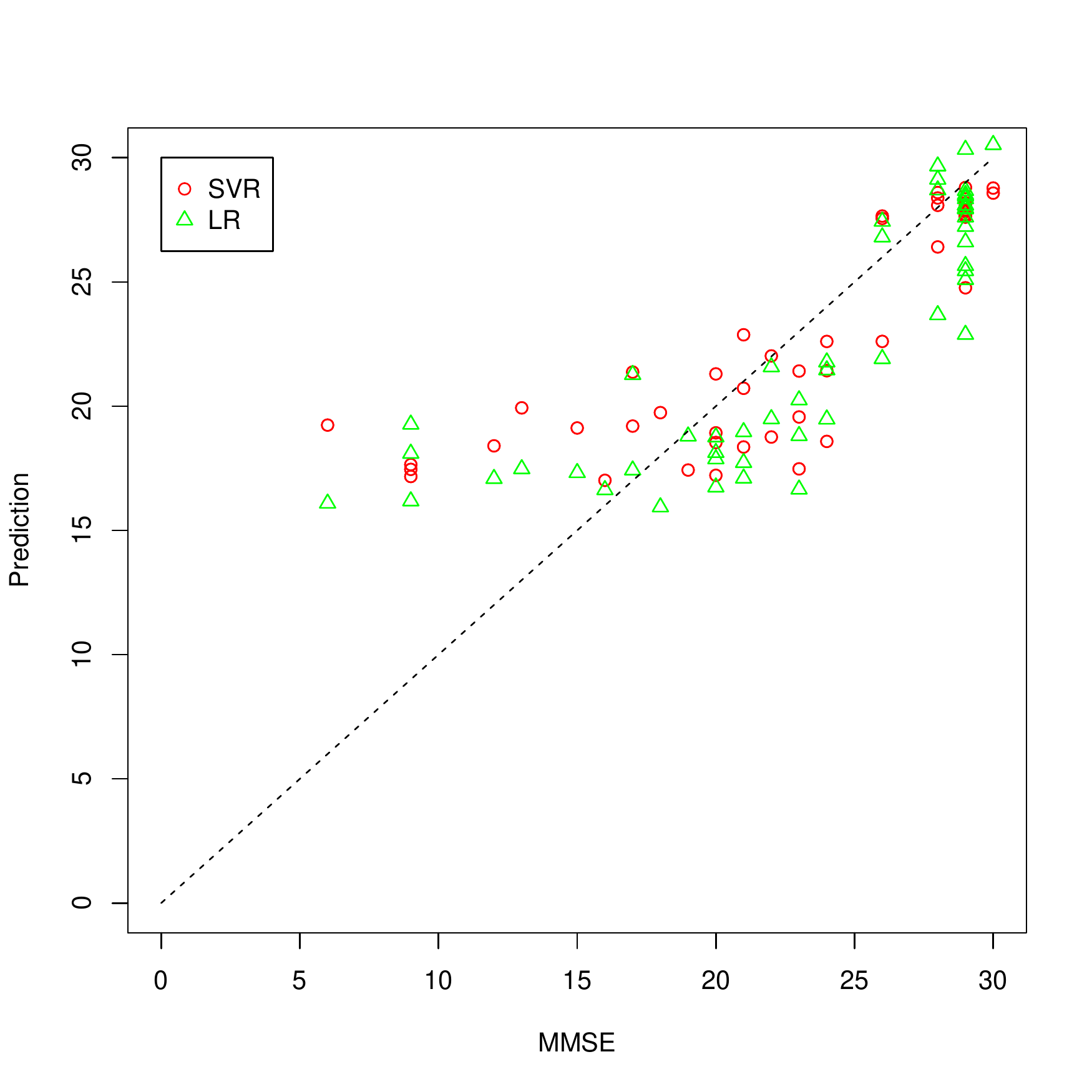}
\caption{Prediction Results with 6 attributes.}
\label{pred6}
\end{figure}

\begin{figure}
\centering
\includegraphics[width=8 cm]{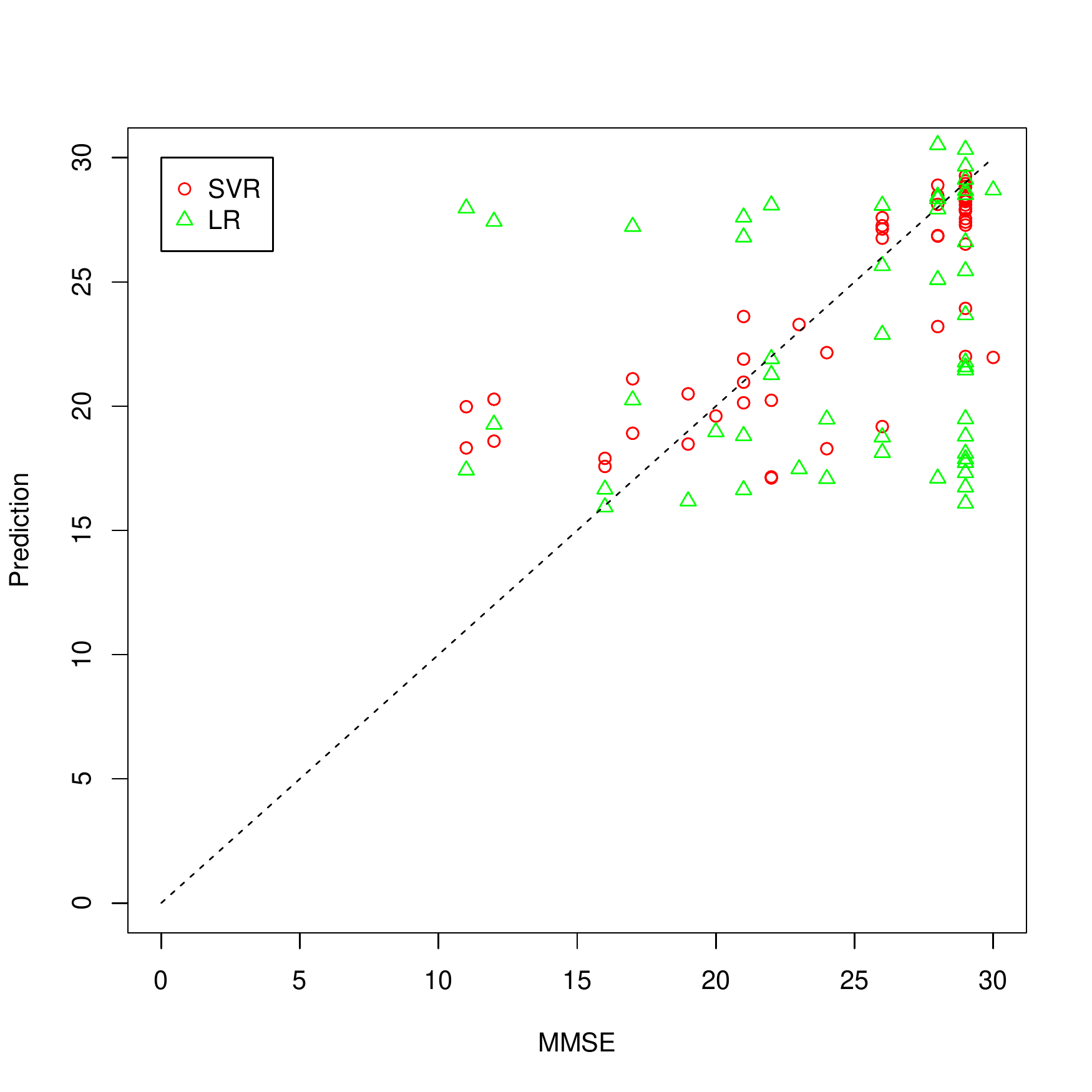}
\caption{Prediction Results with 10 attributes.}
\label{pred10}
\end{figure}

Experimental results are shown in Figure \ref{fttjmi1},\ref{pairs},\ref{pred6},\ref{pred10}. The associations between 168 attributes and MMSE score measured with CE are shown in Figure \ref{fttjmi1}, from which it can be learned that only some attributes of bimanual in-phase are associated with MMSE while those of bimanual unti-phase do not, and that the attributes `Number of taps', `Average of intervals', and `Frequency of taps' of both hands of bimanual in-phase task (\#33--\#35,\#73--\#75 attributes in Figure \ref{fttjmi1}) are mostly associated with MMSE score. The joint distribution of these attributes and MMSE score is plotted in Figure \ref{pairs}. 

In the first experiment, LR and SVR with the above 6 attributes as inputs present high accuracy, as illustrated in Figure \ref{pred6}. Comparison on performance of 100 independent experiments between the two models in terms of MAE and diagnosis accuracy are listed in Table \ref{t:exp1mae}, from which it can be learned that the proposed method achieves good prediction performance (less than 3 points deviation and above 90\% accuracy). 

In the second experiment, we also try to select more attributes as inputs of the ML models based on association strength. This time, the attributes, `Average of contact duration' and `Number of zero crossover points of acceleration' of both hands of bimanual in-phase task (\#28,\#31,\#68,\#71 attributes in Figure \ref{fttjmi1}) are additionally considered because they are associated with MMSE scores stronger than the rest attributes. With these 10 attributes, the ML models present a comparable result with the case with 6 attributes (see Figure \ref{pred10}). It can be learned that a few additional attributes can hardly improve the accuracy of the models. Comparison on the performances of between the two models were also done based on the results of 100 independent experiments in terms of MAE and diagnosis accuracy, as listed in Table \ref{t:exp2mae}.

\section{Discussion}
\label{s:discussion}
In the experiments, we selected the attributes associated with MMSE score with CE instead of other association measures, such as traditional correlation coefficients (CC). In this case, CE makes no assumption on the underlying distribution of the attributes while CC would assume the Gaussanity of the underlying distribution which is violated by the non-Gaussanity of the attributes as shown in Figure \ref{pairs}.

In the experiments, the most associated attributes are actually related with each other by definition. Since all the tasks in the experiments last for the same 15 seconds, it is obvious that ‘Average of interval’ equals to 15 divided by ‘Number of taps’ and that ‘Frequency of taps’ equals to ‘Number of taps’ divided by 15. So they are essentially same thing, despite of possible numerical errors of device software. They can be considered as dementias biomarkers. 

Among many available predictive models, only LR and SVR are adopted in our experiments because the sample sets in our cases are small. Here simple model, like LR with linearity assumption, is preferred rather than complex models (such as ensemble models or neural networks) in case of over-fitting. SVR is also an good choice for such small sample problems due to its theoretical advantage and practical successes. Experimental results show that the performances of LR and SVR are close in every experiment despite SVR are in nonlinear mode after hyper-parameters are tuned for optimal results. This might suggest MMSE prediction problem could be approximately linear. 

\begin{table}
	\centering
	\caption{Performance of LR and SVR in Experiment 1.}
	\begin{tabular}{l|c|c}
		\toprule
		Model&LR&SVR\\
		\midrule
		MAE&2.693&2.548\\
		\hline
		Diagnosis Accurary(\%)&91.7&92.6\\
		\bottomrule
	\end{tabular}
	\label{t:exp1mae}
\end{table}

\begin{table}
	\centering
	\caption{Average Accuracy of LR and SVR in Experiment 2.}
	\begin{tabular}{l|c|c}
		\toprule
		Model&LR&SVR\\
		\midrule
		MAE&2.731&2.566\\
		\hline
		Diagnosis Accuracy(\%)&91.1&91.9\\
		\bottomrule
	\end{tabular}
	\label{t:exp2mae}
\end{table}

Regarding the accuracy of the two models in the experiments, one can easily learn that LR and SVR show high prediction accuracy with less than 3-points deviation in terms of MAE. Following the common clinical practice that a cutoff score for MMSE is used in dementia diagnosis, most dementia patients in our experiment are labeled correctly by our method in spite of less than 3-points deviation. It is reasonable to expect further improvement on prediction accuracy if larger data sets are collected for better models. 

In our ML pipeline, CE plays a vital role on preparing the inputs of the predictive models. As a rigorously defined mathematical concept, CE has its own advantage. CE, a universal measure of statistical independence, can be explained as a quantity of energy exchange or information transmission in real physical or biological systems and therefore the predictive model based on the associations discovered by CE is explainable, which is of vital importance for clinical use.

In this research, the strong associations between certain finger-tapping attributes and MMSE score measured by CE could also has biological meanings, which indicate that hands motor ability and cognitive ability are biologically related with each other and are both syndromes of dementia, more broadly aging. Note that, intuitively, finger-tapping task is so easy to perform that it might relate to only some but not all aspects of mental state measured by MMSE. We suggest further physiological and pathophysiological research on the relation between motor ability and cognitive ability, which could lead to refinement of our technology and better understandings  mechanism of dementia. 

\section{Conclusions}
\label{s:con}
In this paper, we propose a method for predicting MMSE score from finger-tapping measurement with machine learning models. The method is developed with a ML pipeline which comprises of CE based variable selection and predictive models. In variable selection step, based on the data collected from healthy subjects and patients, the associations between certain finger-tapping attributes and MMSE score are measured with CE, and the mostly associated attributes (`Number of taps', `Average of intervals', and `Frequency of taps' of both hands of bimanual in-phase task) are selected. Then the selected attributes are fed into the predictive models (LR and SVR) to predict MMSE score. Experimental results showed that the models such built present good prediction performance. Due to the associations discovered by CE, the predictive model is explainable for clinical use.

The experimental results that certain finger-tapping attributes are associated with MMSE score indicate that motor ability and cognitive ability are associated since the finger-tapping attributes measure motor ability and MMSE measures cognitive ability. Such association may means that both abilities are intrinsically related with each other and that declines of both abilities are syndromes of dementias, or broadly aging. 

\section*{Acknowledgements}
The author thanks Lin Xiaolie and Yin Ying for providing finger-tapping data.

\bibliographystyle{spphys}
\bibliography{bib-mmse}

\begin{thebibliography}{10}
\providecommand{\url}[1]{{#1}}
\providecommand{\urlprefix}{URL }
\expandafter\ifx\csname urlstyle\endcsname\relax
  \providecommand{\doi}[1]{DOI \discretionary{}{}{}#1}\else
  \providecommand{\doi}{DOI \discretionary{}{}{}\begingroup
  \urlstyle{rm}\Url}\fi

\bibitem{un2017}
{United Nations}.
\newblock World population prospects (2017)

\bibitem{who2013}
{World Health Orgnization}.
\newblock Global health estimates summary tables: Deaths by cause, age and sex
  by various regional grouping (2013)

\bibitem{prince2009world}
M.~{Prince}, C.P. {Ferri}.
\newblock World alzheimer report 2009 (2009)

\bibitem{prince2018world}
M.~{Prince}, R.~{Bryce}, C.~{Ferri}.
\newblock World alzheimer report 2011 : The benefits of early diagnosis and
  intervention (2018)

\bibitem{petersen2001practice}
R.C. {Petersen}, J.~{Stevens}, M.~{Ganguli}, E.G. {Tangalos}, J.~{Cummings},
  S.T. {DeKosky}, Neurology \textbf{56}(9), 1133 (2001)

\bibitem{mckhann1984clinical}
G.M. {McKhann}, D.A. {Drachman}, M.F. {Folstein}, R.~{Katzman}, D.L. {Price},
  E.M. {Stadlan}, Neurology \textbf{34}(7), 939 (1984)

\bibitem{1994PracticePF}
Neurology \textbf{44}, 2203  (1994)

\bibitem{doody2001practice}
R.S. {Doody}, J.C. {Stevens}, C.~{Beck}, R.M. {Dubinsky}, J.A. {Kaye},
  L.~{Gwyther}, R.C. {Mohs}, L.J. {Thal}, P.J. {Whitehouse}, S.T. {DeKosky},
  J.L. {Cummings}, Neurology \textbf{56}(9), 1154 (2001)

\bibitem{dubois2007research}
B.~{Dubois}, H.H. {Feldman}, C.~{Jacova}, S.T. {DeKosky},
  P.~{Barberger-Gateau}, J.L. {Cummings}, A.~{Delacourte}, D.~{Galasko},
  S.~{Gauthier}, G.~{Jicha}, K.~{Meguro}, J.~{O'Brien}, F.~{Pasquier},
  P.~{Robert}, M.~{Rossor}, S.~{Salloway}, Y.~{Stern}, P.J. {Visser},
  P.~{Scheltens}, Lancet Neurology \textbf{6}(8), 734 (2007)

\bibitem{albert2011the}
M.S. {Albert}, S.T. {DeKosky}, D.~{Dickson}, B.~{Dubois}, H.H. {Feldman}, N.C.
  {Fox}, A.~{Gamst}, D.M. {Holtzman}, W.J. {Jagust}, R.C. {Petersen}, P.J.
  {Snyder}, M.C. {Carrillo}, C.H. {Phelps}, Alzheimer's \& Dementia
  \textbf{7}(3), 270 (2011)

\bibitem{folstein1975mini}
M.F. {Folstein}, S.E.B. {Folstein}, P.R. {McHugh}, Journal of Psychiatric
  Research \textbf{12}(3), 189 (1975)

\bibitem{tombaugh1992the}
T.N. {Tombaugh}, N.J. {McIntyre}, Journal of the American Geriatrics Society
  \textbf{40}(9), 922 (1992)

\bibitem{kandori2004quantitative}
A.~{Kandori}, M.~{Yokoe}, S.~{Sakoda}, K.~{Abe}, T.~{Miyashita}, H.~{Oe},
  H.~{Naritomi}, K.~{Ogata}, K.~{Tsukada}, Neuroscience Research
  \textbf{49}(2), 253 (2004)

\bibitem{suzumura2016assessment}
S.~{Suzumura}, A.~{Osawa}, T.~{Nagahama}, I.~{Kondo}, Y.~{Sano}, A.~{Kandori},
  Japanese Journal of Comprehensive Rehabilitation Science \textbf{7}, 19
  (2016)

\bibitem{joe2014dependence}
H.~{Joe}, \emph{Dependence Modeling with Copulas} (CRC press, 2014)

\bibitem{nelsen1998an}
R.B. {Nelsen}, \emph{An Introduction to Copulas} (Springer Science \& Business
  Media, 1998)

\bibitem{sklar1959fonctions}
M.~{Sklar}, Publ. Inst. Statist. Univ. Paris \textbf{8}, 229 (1959)

\bibitem{ma2011mutual}
J.~{Ma}, Z.~{Sun}, Tsinghua Science \& Technology \textbf{16}(1), 51 (2011)

\bibitem{cover1991elements}
T.M. {Cover}, J.A. {Thomas}, \emph{Elements of information theory} (John Wiley
  \& Sons, 1991)

\bibitem{kraskov2004estimating}
A.~{Kraskov}, H.~{Stögbauer}, P.~{Grassberger}, Physical Review E
  \textbf{69}(6), 66138 (2004)

\bibitem{smola2004a}
A.J. {Smola}, B.~{Schölkopf}, Statistics and Computing \textbf{14}(3), 199
  (2004)

\bibitem{jian2019variable}
J.~Ma, arXiv preprint arXiv:1910.12389  (2019)

\bibitem{copent}
J.~Ma, \emph{copent: Estimating Copula Entropy and Transfer Entropy} (2021).
\newblock \urlprefix\url{https://CRAN.R-project.org/package=copent}.
\newblock R package version 0.2

\bibitem{Chang2011LIBSVMAL}
C.C. Chang, C.~Lin, ACM Trans. Intell. Syst. Technol. \textbf{2}, 27:1 (2011)

\end{thebibliography}

\end{document}